\documentclass[preprint,journal]{vgtc}       

\ifpdf
  \pdfoutput=1\relax                   
  \usepackage{graphicx}                
  \DeclareGraphicsExtensions{.pdf,.png,.jpg,.jpeg} 
\else
  \usepackage{graphicx}                
  \DeclareGraphicsExtensions{.eps}     
\fi%

\graphicspath{{figures/}{pictures/}{images/}{./}} 

\usepackage{microtype}                 
\PassOptionsToPackage{warn}{textcomp}  
\usepackage{textcomp}                  
\usepackage{mathptmx}                  
\usepackage{times}                     
\usepackage{cite}                      
\usepackage{tabu}                      
\usepackage{booktabs}                  
\usepackage{amsmath}
\usepackage[ruled,vlined]{algorithm2e}
\usepackage{enumitem}
\DeclareMathAlphabet{\mathcal}{OMS}{cmsy}{m}{n} 
\usepackage{xcolor} 

\onlineid{1044}

\vgtccategory{Research}
\vgtcpapertype{algorithm/technique}

\title{ARC: Alignment-based Redirection Controller for Redirected Walking in Complex Environments}

\author{Niall L. Williams, \textit{Student Member, IEEE}, Aniket Bera, \textit{Member, IEEE}, Dinesh Manocha, \textit{Fellow, IEEE}}
\authorfooter{
\item Niall L. Williams, Aniket Bera, and Dinesh Manocha are with the University of Maryland, College Park. E-mail: \{niallw $\vert$ ab $\vert$ dm\}@cs.umd.edu.
}

\shortauthortitle{ARC: Alignment-based Redirection Controller for Redirected Walking}
\shortauthortitle{Niall L. Williams \MakeLowercase{\textit{et al.}}: ARC: Alignment-based Redirection Controller for Redirected Walking in Complex Environments}

\abstract{We present a novel redirected walking controller based on alignment that allows the user to explore large and complex virtual environments, while minimizing the number of collisions with obstacles in the physical environment. 
Our alignment-based redirection controller, ARC, steers the user such that their proximity to obstacles in the physical environment matches the proximity to obstacles in the virtual environment as closely as possible.
To quantify a controller's performance in complex environments, we introduce a new metric, Complexity Ratio (CR), to measure the relative environment complexity and characterize the difference in navigational complexity between the physical and virtual environments.
Through extensive simulation-based experiments, we show that ARC significantly outperforms current state-of-the-art controllers in its ability to steer the user on a collision-free path.
We also show through quantitative and qualitative measures of performance that our controller is robust in complex environments with many obstacles.
Our method is applicable to arbitrary environments and operates without any user input or parameter tweaking, aside from the layout of the environments.
We have implemented our algorithm on the Oculus Quest head-mounted display and evaluated its performance in environments with varying complexity. 
Our project website is available at \href{https://gamma.umd.edu/arc/}{\texttt{https://gamma.umd.edu/arc/}}.
} 


\CCScatlist{ 
 \CCScat{K.6.1}{Management of Computing and Information Systems}%
{Project and People Management}{Life Cycle};
 \CCScat{K.7.m}{The Computing Profession}{Miscellaneous}{Ethics}
}

\teaser{
  \centering
  \includegraphics[width=\linewidth]{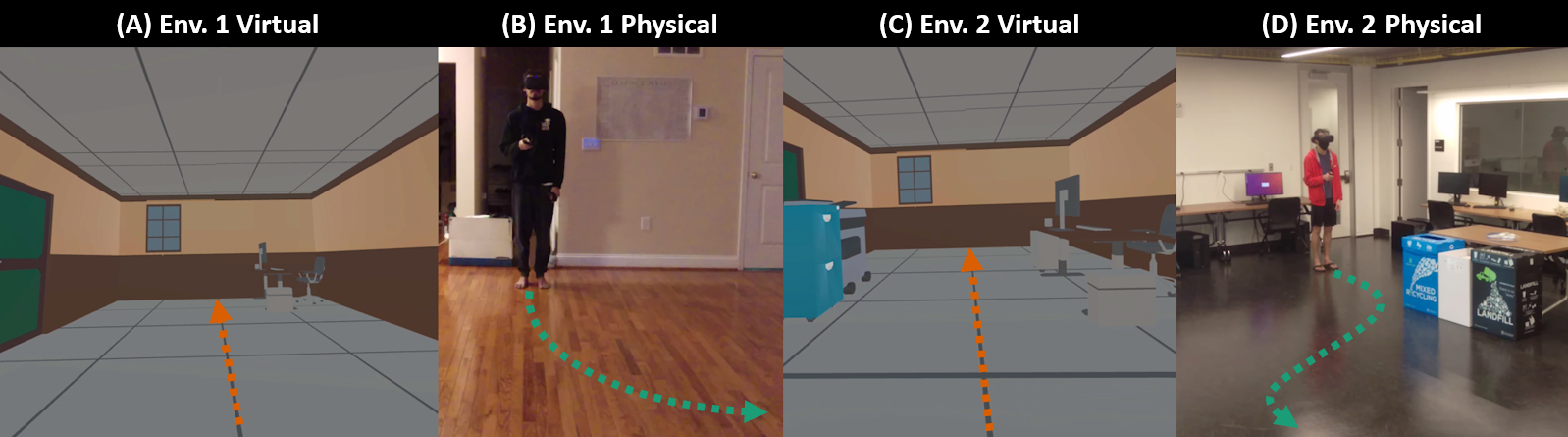}
  \caption{A user being steered with our alignment-based redirection controller (ARC) in two different environments. In Environment 1, the virtual environment (VE) is larger than the physical environment (PE), and there is an obstacle in the northeast corner of the VE. The PE has no obstacles.
  In Environment 2, the VE is larger than the PE, and both have obstacles in different positions.
  \textbf{(A)} The user walks in a straight line forward in the VE. \textbf{(B)} In the PE, the user is steered on a curved path away from the edge of the tracked space, in order to minimize the differences in proximity to obstacles in PE and VE. \textbf{(C)} The user walks in a straight line forward in the VE, with obstacles on either side of the path. \textbf{(D)} The user is steered on a path with multiple curves in the physical space. The user avoids a collision with the obstacle in front of them, and is also steered to minimize the differences in proximity to obstacles in the PE and VE.
  We are able to steer the user along smooth, collision-free trajectories in the PE.
  Our extensive experiments in real-wold and simulation-based experiments show that in simple and complex environments, our approach results in  fewer collisions with obstacles and lower steering rate than current state-of-the-art algorithms for redirected walking.
  }
	\label{fig:teaser}
}

\vgtcinsertpkg

\begin{document}


\firstsection{Introduction}

\maketitle



Exploring virtual environments (VEs) is an integral part of immersive virtual experiences.
Real walking is known to provide benefits to sense of presence \cite{usoh1999walking} and task performance \cite{ruddle2009benefits} that other locomotion interfaces cannot provide.
Using an intuitive locomotion interface like real walking has benefits to all virtual experiences for which travel is crucial, such as virtual house tours and training applications.
Redirected walking (RDW) is a locomotion interface that allows users to naturally explore VEs that are larger than or different from the physical tracked space, while minimizing how often the user collides with obstacles in the physical environment (PE) \cite{razzaque2005redirected}.
RDW works by slowly transforming the VE with rotations or translations such that these transformations are imperceptible while still creating a subtle discrepancy between the user's physical and virtual trajectories.
This discrepancy causes the user to adjust their physical trajectory to counteract the virtual camera motion induced by the virtual transformations, in order to stay on their intended virtual trajectory.
RDW is an appealing locomotion interface because it allows users to explore the VE using real, natural walking.

Although RDW is a good locomotion interface for immersive virtual experiences, it has two main limitations.
First, the amount of redirection that can be applied to steer the user is dependent on how easily the user can consciously perceive the virtual transformations induced by RDW.
The amount of redirection applied is controlled by \textit{gains}, which determine the magnitude (intensity) of rotations and translations applied to the VE.
The gains that transform the VE the most, while still remaining imperceptible to the user, are the \textit{perceptual threshold} gains \cite{steinicke2009estimation}.
The intensity of a gain corresponds to the amount of deviation between the user's physical and virtual paths, and thus the amount that the user is steered in the PE.
Gains with high intensity result in more redirection of the physical user at the cost of larger VE transformations, which can be more easily detected and can cause simulator sickness \cite{nilsson201815}.

The second limitation is that the effectiveness of RDW at minimizing the number of collisions with physical obstacles depends on the \textit{relative complexity} of the physical and virtual environments.
Navigation through an environment becomes more difficult when the environment is populated with more obstacles because the user has fewer options for possible collision-free routes that allow them to avoid collisions with obstacles when making any movements.
Thus, the complexity of an environment can be described in terms of the density of the obstacles in the environment.
For example, the empty VE used by Bachmann et al. \cite{bachmann2019multi} would be considered low complexity, while the maze-like VE with multiple branching paths used by Nescher et al. \cite{nescher2014planning} would be considered high complexity.
In virtual reality (VR), the user navigates through a virtual and physical environment \textit{at the same time}.
A movement in one environment is paired with a movement in the other.
Therefore, the navigation problem becomes harder, since all movements must consider the obstacles in the VE as well as the PE.
If the complexities (density and layout of obstacles) of the PE and VE are similar, avoiding collisions is easier, since a movement that yields a particular result (collision or no collision) in one environment is likely to result in the same outcome in the other environment.
However, if the complexities of the PE and VE are very different, navigation is harder, because a movement in one environment will likely lead to a movement in the other environment with a different outcome.


In the context of VR, a \textit{redirection controller} determines the amount of redirection to apply, given the user's position in the physical and virtual environments \cite{nilsson201815}.
At each frame, the controller decides the level (intensity) of gains to apply in order to rotate or translate the VE and alter the user's physical trajectory.
The controller uses heuristics or optimization in conjunction with information about the PE and/or VE in order to determine the level of gains to apply.
\textit{Reactive} controllers make decisions on how to steer the user based on the instantaneous state of the system, while \textit{predictive} controllers make decisions based on predictions about the user's future trajectory.
Reactive controllers typically do not consider the VE when setting gains, and the VE is often abstracted away by using an unbounded, empty environment.
This design decision allows reactive controllers to be simpler and remain relatively effective without requiring much additional information, at the cost of worse performance than predictive controllers in some environments.
On the other hand, predictive algorithms often use information from the VE to make predictions about the user's virtual trajectory. 
From these predictions, these algorithms are able to perform better than reactive algorithms by applying gains that are more suited to the user's environments and trajectory \cite{nescher2014planning, zmuda2013optimizing}.
Predictive algorithms rely on accurate predictions however, so they usually do not perform well if it is difficult to forecast the user's movement.

At a high level, \textit{alignment} can be defined as a state in which the user's physical and virtual configurations match.
It was first formally studied by Thomas et al. \cite{thomas2020reactive, thomas2020towards} with the goal of allowing the user to interact with the PE when they are within some predefined region of the VE, to enable haptics.
Prior to the work of Thomas et al., Zmuda et al. \cite{zmuda2013optimizing} developed a controller that used a core idea of alignment; their controller was allowed to place the user near a physical obstacle if its position relative to the physical user matched the position of a virtual obstacle relative to the virtual user.
Alignment provides a simple way to consider both the physical \textit{and} virtual environment when steering the user, which is important for developing effective controllers.

\noindent {\bf Main Contributions:} In this paper, we present a novel alignment-based redirection controller (ARC) for locomotion in VR.
ARC is a redirection controller that applies RDW gains to steer the user such that the user's proximity to obstacles in the PE matches their proximity to obstacles in the VE as closely as possible.
Our controller is able to steer users through physical and virtual environments that have different relative complexities and makes no assumptions about the distribution of the obstacles in each environment.
In these complex environments, ARC achieves a lower number of collisions with physical obstacles when compared to current state-of-the-art controllers.
Furthermore, ARC achieves this lower number of collisions while also redirecting the user using \textit{less intense} gain than other controllers, which reduces the likelihood that users experience simulator sickness during locomotion \cite{nilsson201815}.
We conduct extensive experiments in varied environments, using many different performance metrics, and find that ARC consistently outperforms existing state-of-the-art algorithms.
The main contributions included in this paper are:
\setlist{}
\begin{itemize}[]
    \item A novel alignment-based redirection controller that can function in arbitrary environments, without requiring any information from the application except for a map of the PE and VE and the obstacles in each environment. Benefits of our algorithm include:
    \begin{itemize}
        \item Significantly fewer collisions in PE/VE pairs with similar complexities \textit{and} in PE/VE pairs with very different complexities.
        \item A lower steering rate, which helps avoid simulator sickness and increases the usability of the system for users with high sensitivity to redirection.
    \end{itemize}
    \item A novel metric, Complexity Ratio (CR), for measuring and comparing the complexity of PE/VE pairs in the context of VR navigation. Using CR, we can directly assess a controller's performance in different environments, which allows us to compare redirection controllers easily.
    \item Extensive simulation-based evaluation of ARC compared to current state-of-the-art controllers. From our experiments, we conclude that alignment is an effective tool for minimizing collisions with physical obstacles when steering a user with RDW in simple and complex environments. We also show our controller working in a proof of concept implementation on the Oculus Quest.
\end{itemize}

\section{Background} \label{background}
Redirected walking works by imperceptibly transforming the VE around a user such that they adjust their physical trajectory to compensate for the VE transformations and remain on their intended virtual trajectory \cite{razzaque2005redirected}.
The magnitude (or intensity) of the VE transformations is determined by \textit{gains}.
Razzaque et al. \cite{razzaque2005redirected} defined three gains, rotation, translation, and curvature, for rotating or translating the VE depending on the user's movement.
Rotation gains rotate the VE around the user as they turn in place, which results in virtual rotations that are larger or smaller than the corresponding physical rotations, depending on the direction of the VE rotation relative to the physical rotation.
Translation gains translate the VE forward or backward as the user walks in a straight line, which results in their virtual displacement being different from their physical displacement, depending on the direction of the VE translation.
Curvature gains steer the user on a curved physical path by slowly rotating the VE around the user as they walk on a straight virtual path.
The direction in which the user is steered is determined by the direction that the VE is rotated.
Most research in redirected walking aims to either understand the perceptual limits of redirection or develop RDW controllers that minimize the number of collisions a user experiences during locomotion.
An overview of different RDW methods is given in \cite{nilsson201815}.

\subsection{Perceptual Thresholds}
The amount of redirection that can be applied before users notice the redirection is determined by perceptual thresholds.
Perceptual thresholds are important to consider since strong redirection can induce simulator sickness \cite{nilsson201815} and break the user's feeling of presence in a virtual experience \cite{suma2012taxonomy}.
There has been considerable research into measuring the perceptual thresholds of each RDW gain, but there is no general consensus when it comes to selecting these thresholds \cite{nilsson201815}.

The first comprehensive study of RDW thresholds was performed by Steinicke et al. \cite{steinicke2009estimation}.
Many researchers have since expanded on threshold estimation by reproducing results and measuring thresholds under different conditions.
A study by Grechkin et al. \cite{grechkin2016revisiting} determined that translation and curvature gains can be applied simultaneously without altering either gain's perceptual thresholds.
Neth et al. \cite{neth2012velocity} showed that a user's curvature gain detection thresholds are dependent on his or her walking speed.
Williams et al. \cite{williams2019estimation} reproduced the results found by Steinicke et al. \cite{steinicke2009estimation} and demonstrated that users' perceptual thresholds could vary depending on their gender, the field of view, and the presence of distractors in the VE.
Hutton et al. \cite{hutton2018individualized} suggest that perceptual thresholds can differ greatly between different people, which may explain the different threshold values reported in prior literature.
Thus, a system using some commonly-accepted threshold values (such as those measured in \cite{steinicke2009estimation}) may apply appropriate gains for most users. However, gains could still be too high for some users, which could induce sickness and make the user uncomfortable.
All this suggests that there are still many open problems with respect to accurately measuring a person's perceptual thresholds.
A recent review of studies that measured perceptual thresholds can be found in \cite{langbehn2018redirected}.

\subsection{Redirected Walking Controllers}
A redirection controller is an algorithm that decides which gains to apply at each frame in order to minimize the number of collisions the user incurs in the PE \cite{nilsson201815}.
While a controller's goal is to minimize the number of collisions, it is important to note that a controller cannot guarantee a collision-free trajectory in all circumstances.
A controller's effectiveness depends on the configuration of the physical and virtual environments (environment dimensions and size and location of obstacles), the virtual path traveled, and the user's perceptual thresholds.


Controllers fall into three categories: scripted, reactive, and predictive \cite{nilsson201815}.
\textit{Scripted controllers} steer the user as they follow a virtual path pre-determined by the system developers.
Scripted controllers are effective at reducing the number of collisions but impose tight restrictions on the VE.
These controllers can perform very poorly if the user deviates from the pre-determined virtual path.
Work studying scripted controllers includes the development of steering algorithms based on change blindness \cite{suma2011leveraging} and overlapping virtual spaces \cite{suma2012impossible}.

\textit{Reactive controllers} steer the user based on information available from the user's previous movements and current state.
These controllers are designed to work in a wide variety of PEs and VEs since they do not make assumptions about the user's future path.
Reactive controllers fall short at achieving maximal collision avoidance since they do not use all the information available to the system.
Razzaque \cite{razzaque2005redirected} proposed three reactive algorithms for RDW: steer-to-center (S2C), steer-to-orbit, and steer-to-multiple-targets.
Steer-to-center constantly redirects the user towards the center of the physical environment.
Steer-to-orbit steers the user along a circular path that orbits the center of the PE.
Steer-to-multiple-targets steers the user to one of multiple pre-determined physical goal positions, depending on the user's position in the PE.
Despite being one of the first controllers ever, S2C has regularly outperformed other algorithms in a variety of environments \cite{hodgson2013comparing, azmandian2015physical}.
Additionally, steer-to-orbit performs well when the user walks on long, straight virtual paths \cite{hodgson2013comparing}.
However, more recent algorithms have performed as well as, or better than, S2C.
Strauss et al. introduced a controller trained by reinforcement learning that outperformed S2C in simulated trials and performed as well as S2C in user trials \cite{strauss2020steering}.
Chang et al. \cite{chang2019redirection} and Lee et al. \cite{lee2019real} have also recently used reinforcement learning to train RDW controllers.
Thomas et al. \cite{thomas2019general} and Bachmann et al. \cite{bachmann2019multi} simultaneously introduced controllers based on artificial potential fields that outperformed S2C in non-convex and multi-user environments.

\textit{Predictive controllers} make predictions about the user's intended virtual path and steer them accordingly.
Predictive controllers can be effective since they tend to use most of the information available to the system.
However, their performance relies partially on the accuracy of their predictions.
Nescher et al. \cite{nescher2014planning} and Zmuda et al. \cite{zmuda2013optimizing} developed predictive controllers that were outperformed S2C, while Dong et al. improved upon the potential field-based controllers by incorporating trajectory prediction into the controller \cite{dong2020dynamic}.


In addition to determining the gains to apply at each frame, redirection controllers have a resetter component.
When the user gets too close to a physical obstacle, the system initiates a reset maneuver in order to reorient the user away from the nearby obstacle.
This reset maneuver is counted as a collision.
The specific reset policy employed depends on the controller, but one popular resetting technique is the 2:1 reset \cite{williams2007exploring}, wherein the magnitude of a user's physical rotations is doubled, so a $180^\circ$ physical turn yields a $360^\circ$ virtual turn.
Another effective reset technique is distractors, which are elements in the VE that capture the user's attention to reorient them \cite{peck2009evaluation, peck2010improved, peck2011evaluation}.
Other reset techniques may be specific to the RDW controller, such as the reset-to-gradient technique seen in potential field controllers \cite{thomas2019general,bachmann2019multi}.

Evaluation metrics for RDW controllers can depend on the experimental setup.
The majority of all studies use the number of collisions as one performance metric.
Other common metrics include the average virtual distance walked between collisions, the mean steering rate, and user performance at a virtual task.
Since the success of an RDW controller depends on the environments and the path traveled, it can be difficult to compare algorithms using only performance metrics.
Thus, it is common for researchers to test not only their new controller but also the state-of-the-art controllers in the same environments. 

\subsection{Environment Complexity Metrics}
Measuring the effect of environment complexity on task performance is useful for understanding the interactions between an agent and its environment.
This measurement enables us to understand and predict how an agent will perform at a task in an environment, which allows us to change the environment design or improve our algorithms accordingly.

VR researchers have studied how a user's ability to complete a task in an environment depends on the environment's complexity \cite{bowman1998methodology, bowman1999maintaining, ragan2015effects}.
While those studies are useful for understanding the interactions between environment complexity and task performance, they did not quantify the environment complexity with precise metrics.
This makes it difficult to generalize their results and makes it harder to predict how users will perform in unstudied environments.
Researchers in robot navigation have developed metrics to quantify environment complexity \cite{anderson2007proposed, shell2003human, crandall2003towards}.
These metrics allow researchers to group and classify environments by complexity and directly compare the performance of different algorithms in different groups of environments.
\section{Redirection by Alignment} \label{alignment}


The concept of alignment in the context of redirected walking was first formally studied by Thomas et al. \cite{thomas2020reactive, thomas2020towards}.
However, Zmuda et al. \cite{zmuda2013optimizing} used key elements of alignment prior to the work of Thomas et al.
Additionally, Simeone et al. \cite{simeone2020space} recently introduced a locomotion technique that is similar to alignment in that it aims to match the VE and the PE, but it does so by overtly manipulating the VE in real time.
In this section, we define our notion of alignment and provide the details of our general alignment-based redirection controller.

\subsection{Definitions and Background}
\subsubsection{Alignment}
\label{alignment-definition}

A redirection controller takes as input the current position and orientation of the user in the PE and the VE.
We define the configuration of the VR system as the user's position and orientation in the PE and VE.
\textit{Alignment} is a configuration in which the user's physical state matches their virtual state.
When this configuration is achieved, we say that the system (or user) is \textit{aligned}.
In this paper, we are concerned with steering users on collision-free paths in the PE and VE at the same time.
How close a user is to incurring a collision can be described by the distance to obstacles around them, i.e. their proximity to obstacles.
Thus, we describe the user's \textit{state} in an environment by their proximity to obstacles in the environment.

We assume that the user travels on a collision-free path in the VE.
We associate a proximity function along each point on this path.
This proximity function tells us how close the point is to obstacles in the environment.
We want to define a proximity function that can be formulated for paths in the PE and VE, to be used by our redirection controller to compute collision-free paths to steer the user on.

Let $d(p, \theta)$ be the distance to the closest obstacle in the direction $\theta$ originating from a location $p = (x,y)$ in an environment.
This distance can be computed using simple ray-intersection queries.
Let $S = \{\theta_1, \theta_2, ..., \theta_k\}$ be a set of $k$ discrete directions in the range $[0, 2\pi)$.
We define the proximity function, $Prox(p)$, at a point $p$ to be the sum of distances to obstacles in each direction $\theta_i \in S$:
\begin{equation}
    Prox(p) = \sum_{i=1}^k{d(p, \theta_i)}.
\end{equation}
To compare proximity values $Prox(p_{phys})$ and $Prox(p_{virt})$ for a point $p_{phys}$ in the PE and a point $p_{virt}$ in the VE, we \textit{cannot} simply compute $Prox(p_{phys}) - Prox(p_{virt})$.
If we did, it would be possible to get a value of 0, implying perfect alignment, for positions in which the physical and virtual user are \textit{not} actually perfectly aligned.
For example, this can happen when $d(p_{phys}, \theta)-d(p_{virt}, \theta)=-1\cdot (d(p_{phys}, \theta+\pi)-d(p_{virt}, \theta+\pi))$ for all $\theta \in [0,2\pi)$.

To resolve this problem and have a meaningful notion of what it means to compare values of $Prox(p)$, we can instead sum the absolute value of the differences in distance over all $\theta_i$:
\begin{equation} \label{eqn:cont_difference}
    dist(Prox(p_{phys}), Prox(p_{virt})) = 
        \sum_{i=1}^{k}{| d(p_{phys}, \theta_i) - d(p_{virt}, \theta_i) |}.
\end{equation}
Note that we use the same set of directions $S$ for the PE and the VE.
Computing this difference is too computationally expensive for large values of $k$, so in our implementation we approximate this value by computing the difference in distances in three directions around the point ($k=3$).

\begin{figure}[t]
    \centering
    \includegraphics[width=.47\textwidth]{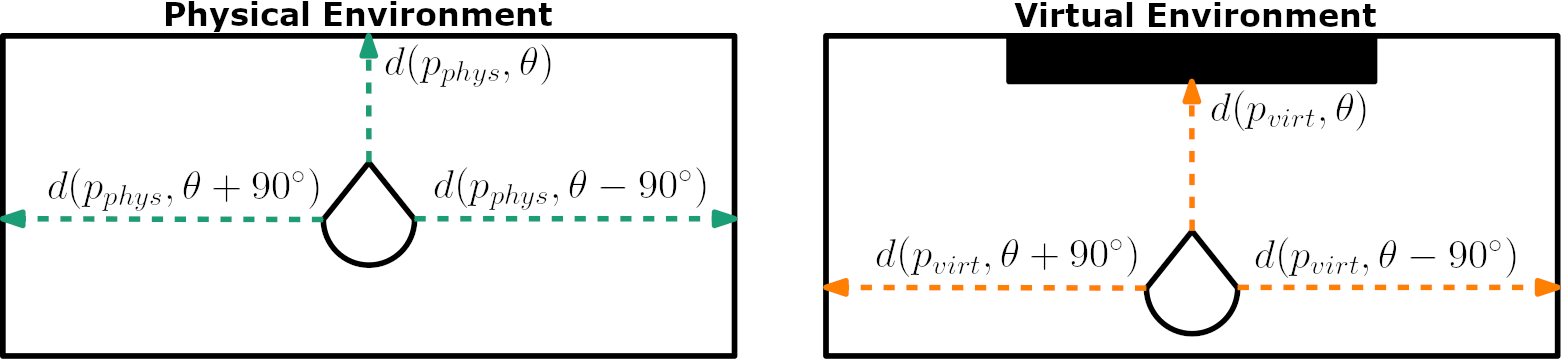}
    \caption{Visualization of the three values from the PE and three values from the VE that constitute a user's state.}
    \label{fig:state}
\end{figure}

Now we can define the physical and virtual states at time $t$ which we use in our rediretion controller:
\begin{equation}
    \begin{aligned}
    q_t^{phys} &= \{ d(p_{phys}, \theta_{phys}), 
                         d(p_{phys}, \theta_{phys}+90^\circ), 
                         d(p_{phys}, \theta_{phys}-90^\circ) \}, \\ 
    q_t^{virt} &= \{ d(p_{virt}, \theta_{virt}), 
                        d(p_{virt}, \theta_{virt}+90^\circ), 
                        d(p_{virt}, \theta_{virt}-90^\circ) \}.
    \end{aligned}
\end{equation}
Here, $p_{phys}$ and $p_{virt}$ are the user's positions in the physical and virtual environments, respectively.
Similarly, $\theta_{phys}$ and $\theta_{virt}$ are the user's headings in the physical and virtual environments, respectively.
Given the user's physical and virtual states, we define the state of the system at time $t$ as the union of their physical and virtual states at time $t$:
\begin{equation}
\label{eqn:state}
    Q_t = \{ q_t^{phys}, q_t^{virt} \}.
\end{equation}
This definition of state is illustrated in \autoref{fig:state}.

Given $Q_t$, we can measure the alignment of the state, $A(Q_t)$ by computing the discretized version of \autoref{eqn:cont_difference}:
\begin{equation}
    A(Q_t) = dist(q_t^{phys}, q_t^{virt}),
\end{equation}
where $dist(q_t^{phys}, q_t^{virt})$ is defined as the sum of the absolute values of the differences between the distances to obstacles in the PE and VE:
\begin{equation}
    \begin{aligned}
    dist(q_t^{phys}, q_t^{virt}) &= 
            | d(p_{phys}, \theta_{phys}) - d(p_{virt}, \theta_{virt}) | \\
            & + \ | d(p_{phys}, \theta_{phys}+90^\circ) - d(p_{virt}, \theta_{virt}+90^\circ) | \\
            & + \ | d(p_{phys}, \theta_{phys}-90^\circ) - d(p_{virt}, \theta_{virt}-90^\circ) |.
    \end{aligned}
\end{equation}

The more similar a user's physical and virtual states are, the closer $A(Q_t)$ will be to 0.
Conversely, a physical and virtual state that are very different will yield a larger value for $A(Q_t)$.
We reiterate that one can define $A(Q_t)$ and $dist(q_t^{phys}, q_t^{virt})$ differently from how we have defined them.
A different definition corresponds to a different notion of what it means for a system to be aligned.
Our controller is concerned with avoiding collisions in the PE, so proximity to obstacles was an appropriate way to define $A(Q_t)$ and $dist(q_t^{phys}, q_t^{virt})$.

With traditional RDW controllers, the goal of the system is to steer the user away from obstacles in the PE.
With an alignment-based controller, the goal is to steer the user to a physical state that most closely matches the virtual state.
In general, the VE will differ greatly from the PE, so it is common that a particular virtual state does not have a corresponding physical state with which it aligns perfectly.
Thus, at any given instance, an alignment-based redirection controller aims to minimize the difference between the physical and virtual states and does \textit{not} necessarily aim to perfectly align the two.
If the global minimum yields perfect alignment, then an alignment-based controller should eventually reach this configuration.
If it were possible to keep a user aligned at all times, the user would never encounter collisions while exploring a VE, and an alignment-based RDW controller could provide an optimal solution to the problem of RDW.

\subsubsection{Environment Complexity}
The complexity of an environment is dependent on the task to be completed in the environment \cite{crandall2003towards}.
For the purposes of locomotion, we define complexity as the ease with which a user can reach a goal destination without colliding with any obstacles in the environment.
As the environment becomes populated with more obstacles, this becomes more difficult, and so the complexity of the environment increases.

Let $C(p)$ be the shortest distance between a point $p$ and the closest obstacle in an environment $E$.
We define the complexity of $E$ as the average value of $C(p)$ over all points in $E$:
\begin{equation}
\label{eqn:env-complexity}
    C(E) = \frac{1}{|P|}\sum_{p\in P}^{}C(p),
\end{equation}
where $P$ is the set of all points $p$ in $E$.
When there is a lot of open space in the environment (low obstacle density), $C(E)$ will be large; $C(E)$ approaches $0$ as the amount of open space in the environment decreases (high obstacle density).

Since VR locomotion depends on a physical \textit{and} virtual environment, we must relate the environments' complexity measures together.
To do this, we compute the \textit{complexity ratio} (CR) of the virtual environment $E_{virt}$ and physical environment $E_{phys}$:
\begin{equation}
    \text{CR} = \frac{E_{phys}}{E_{virt}}.
\end{equation}

This definition of environment complexity gives us an intuitive way to describe how easy it is to locomote through an environment, and CR tells us how similar the complexities of two environments are.
In VR, it is common for the PE to have a lower obstacle density than the VE, \textit{i.e.} $C(E_{phys}) > C(E_{virt})$.
A higher value for CR corresponds to a greater disparity in the complexity of the PE and VE, which implies that collision-free VR navigation is more difficult.
This definition for CR is formulated on a continuous domain, which makes it difficult to compute exactly.
To simplify the computation, we discretize the equation by sampling a point every $0.5$ meters in the environment.
\color{black}

\subsection{Alignment-based Redirection Controller}
\label{subsec:alignment-controller}


\subsubsection{Redirection Heuristic}
In this section, we provide details on how our alignment-based redirection controller (ARC) uses alignment to determine the RDW gains to apply at each frame.
Note that ARC assumes that the user travels on a collision-free virtual trajectory in the direction of their heading (they do not walk backwards or side-to-side).
Since the user's positions and orientations in the environments are known at all times through the tracking information, ARC can compute the $A(Q_t)$ on every frame according to the equations defined in \autoref{alignment-definition}.
If $A(Q_t) = 0$, no redirection is applied.
If $A(Q_t) \neq 0$, ARC uses the following heuristics to set the redirection gains, depending on the user's current movement.

If the user is translating, the translation gain $g_{t}$ is set to be:
\begin{equation}
    g_{t} = \text{clamp}\left(\frac{d(p_{phys}, \theta_{phys})}{d(p_{virt}, \theta_{virt})},\ minTransGain,\ maxTransGain\right),
\end{equation}
where $minTransGain = 0.86$ and $maxTransGain = 1.26$.
The clamp$(x,minVal,maxVal)$ function returns $x$ if $minVal \leq x \leq maxVal$, and returns $minVal$ if $x < minVal$ or $maxVal$ if $x > maxVal$.
This heuristic for the translation gain speeds up the user's physical walking speed relative to their virtual walking speed if there is more open space in front of the physical user than there is in front of the virtual user.
If there is more space in front of the virtual user than there is in front of the physical user, the user's physical walking speed decreased relative to their virtual walking speed.
We set $g_{t}$ equal to the ratio of the distances, bounded by previous measured perceptual thresholds, so that the translation gain changes gradually, which increases user comfort.

If the user is undergoing a translation motion, we need to set the curvature gain $g_{c}$.
First, we determine which of the spaces to the left and right of the physical user is more dissimilar to its virtual counterpart:
\begin{equation}
    \begin{aligned}
    misalignLeft &= d(p_{phys}, \theta_{phys}+90^\circ) - d(p_{virt}, \theta_{virt}+90^\circ), \\
    misalignRight &= d(p_{phys}, \theta_{phys}-90^\circ) - d(p_{virt}, \theta_{virt}-90^\circ).
    \end{aligned}
\end{equation}
If $misalignLeft > misalignRight$, we want to steer the user to the left in order to minimize the misalignment.
To do this, we set $g_{c}$ as follows:
\begin{equation}\label{eqn:leftCurveGain}
    \begin{aligned}
    scalingFactor &= min(1, misalignLeft), \\
    g_{c} &= min(1, scalingFactor \times maxCurvatureRadius),
    \end{aligned}
\end{equation}
where $maxCurvatureRadius = 7.5m$.
If we instead want to steer the user to the right in order to minimize the misalignment (i.e. when $misalignRight > misalignLeft$), we set $g_{c}$ in a manner similar to \autoref{eqn:leftCurveGain}, but we exchange $misalignLeft$ for $misalignRight$ in the $scalingFactor$ computation.
The sign of the curvature gain must also be set appropriately to steer the user in the desired direction.
With this heuristic, we set the curvature gain proportional to the misalignment on the left or right of the user, depending on which is larger.
The gain is bounded by the maximum curvature gain of radius $7.5m$ to reduce the chance that the user feels simulator sickness during redirection.

If the user is rotating, we set the gain according to the user's distance to objects in front of \textit{and} on both sides of the user.
Our heuristic for $g_{t}$ only considers the distance to objects in front of the user since translation gains only alter the forward and backwards displacement of the user.
The heuristic for $g_{c}$ only considers the distances to objects on either side of the user since curvature gains only steer the user to the left or right when walking on a straight virtual path.
However, our heuristic for the rotation gain considers all three distances in order to accurately describe the user's orientation.
Orientation is a function of all $360^\circ$ around the observer, so the most accurate measurement of orientational alignment would compare distances in all $360^\circ$ directions.
That degree of detail is not necessary, since $d(p,\theta)$ and $d(p,\theta+ \Delta\theta)$ will produce very similar values for most positions $p$ in an environment for small $\Delta\theta$.
Furthermore, sampling distances in all directions around the observer is too computationally expensive to run in real-time, which is a requirement for reactive RDW controllers.

To set the rotation gain $g_{r}$, we check if the direction that the user is turning increases or decreases their rotational alignment.
To compute this, we first compute the user's rotational alignment for the current and previous frames:
\begin{equation}\label{}
    \begin{aligned}
    curRotaAlignment &= dist(q_{t}^{physical}, q_{t}^{virtual}), \\
    prevRotaAlignment &= dist(q_{t-1}^{physical}, q_{t-1}^{virtual}).
    \end{aligned}
\end{equation}
Then, we set $g_{r}$ based on whether the current rotational alignment is better or worse than the previous rotational alignment:
\begin{equation}
    g_{r} = 
    \begin{cases}
        minRotaGain & prevRotaAlignment < curRotaAlignment, \\
        maxRotaGain & prevRotaAlignment > curRotaAlignment, \\
        1 & \text{otherwise}.
    \end{cases}
\end{equation}
Here, $minRotaGain = 0.67$ and $maxRotaGain = 1.24$.
The rotation gain is smoothed by linearly interpolating $g_r$ between frames with a weighting of 0.125 on the previous frame's gain.
The idea behind this heuristic is that we want to speed up the user's rotation when they are turning in a direction that improves their rotational alignment, and slow down their rotation when they are turning in a direction that worsens their rotational alignment.

\subsubsection{Resetting Heuristic}
\label{subsubsec:reset}
Since we cannot guarantee that the user will travel on a collision-free physical path, our alignment-based controller needs a resetting policy to reorient the user when they are about to collide with a physical obstacle.
To ensure that the user does not actually walk into any obstacles, a reset is triggered when the user comes within $0.7m$ of an obstacle.
Our reset policy reorients the user such that they face the direction in the PE for which the distance to the closest physical obstacle in the user's physical heading direction most closely matches the distance to the closest virtual obstacle in the user's virtual heading direction.

\begin{figure}[]
    \centering
    \includegraphics[width=0.45\textwidth]{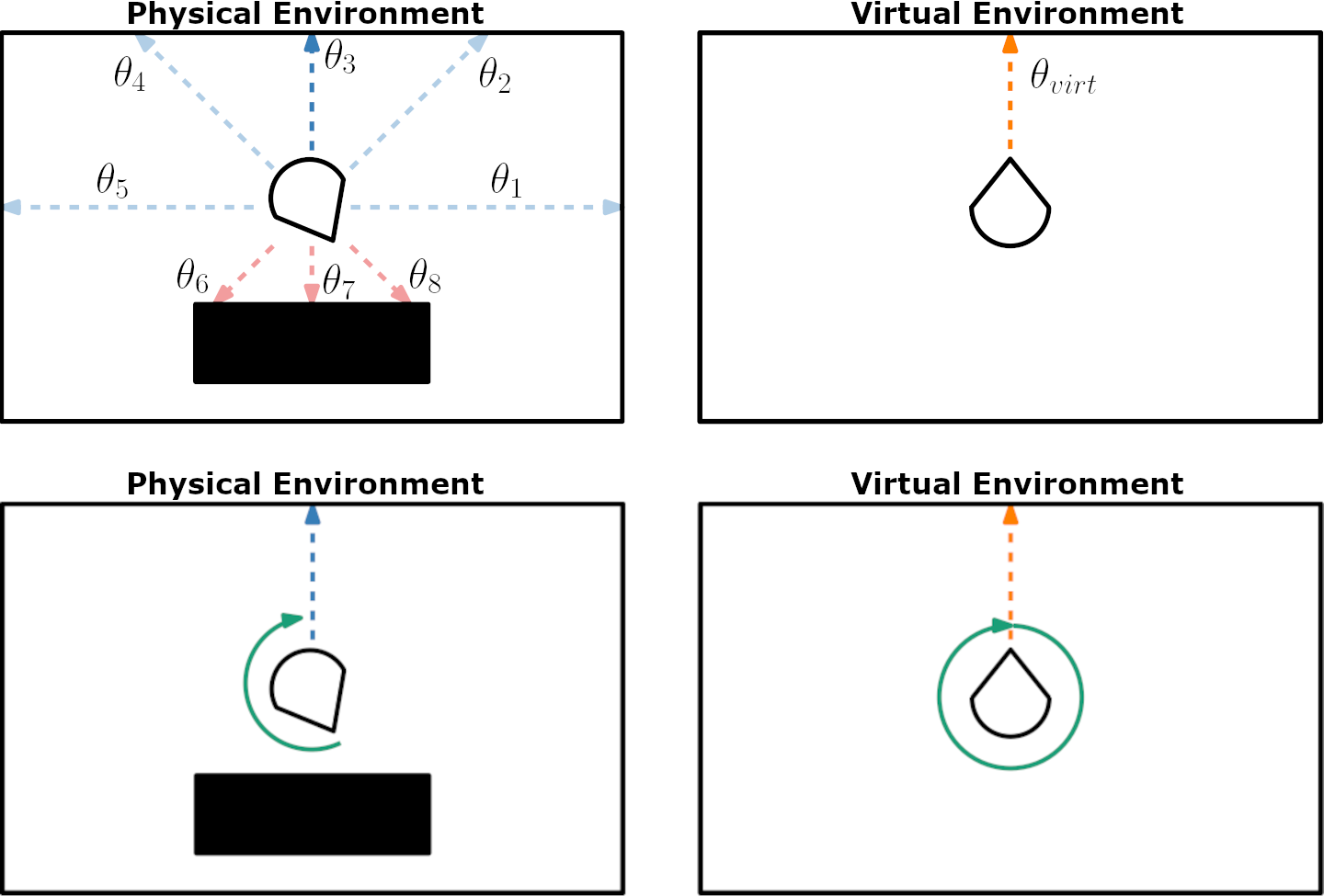}
    \caption{A visualization of the two steps involved in resetting. The top row shows the process of selecting the best direction for resetting. In this example, $\theta_{reset}$ is chosen to be $\theta_3$. To reduce visual clutter, we only show eight of the twenty sampled directions. The bottom row shows the user to turning to face the best direction.}\label{fig:reset}
\end{figure}

When a reset is triggered, let $p_{phys}$ and $p_{virt}$ be the user's physical and virtual positions, respectively, and let $\theta_{phys}$ and $\theta_{virt}$ be their physical and virtual headings, respectively.
First, we sample 20 equally-spaced directions $\{ \theta_1, \theta_2, ..., \theta_{20} \}$ on the unit circle centered at $p_{phys}$.
For each $\theta_i$, we compute the distance to the closest physical obstacle in that direction as $d(p_{phys}, \theta_i)$ to produce 20 distances $\{ d_1, d_2, ..., d_{20} \}$.
The direction the user will face after the reset is complete, denoted $\theta_{reset}$, is the $\theta_i$ for which the corresponding distance $d_i$ most closely matches the distance $d(p_{virt}, \theta_{virt})$.
This value $\theta_{reset}$ is subject to two constraints.
First, $\theta_{reset}$ must face away from the obstacle the user is about to walk into:
\begin{equation}
    dot(\theta_{reset}, obstacleNormal) > 0,
\end{equation}
where $obstacleNormal$ is the normal of the closest face of the obstacle that triggered the reset.
The second constraint is that $d(p_{phys}, \theta_{reset}) \geq d(p_{virt}, \theta_{virt})$.
If this second constraint cannot be satisfied by any of the $\theta_i$ that also satisfy the first constraint, $\theta_{reset}$ is set to the direction that minimizes the difference between $d(p_{phys}, \theta_{reset})$ and $d(p_{virt}, \theta_{virt})$ and satisfies the first constraint. 
Please refer to \autoref{fig:reset} for a visual explanation of our resetting policy.

Once $\theta_{reset}$ is computed, we turn the user to face that direction.
The user is instructed to turn in place until their heading is the same as $\theta_{reset}$, while the virtual turn is scaled up to be a $360^\circ$ turn.
In order to minimize the amount of rotational distortion required for the reset, the user turns in the direction of the larger of the two angles between $\theta_{phys}$ and $\theta_{reset}$.
This method for resetting is inspired by the reset method used by Bachmann et al. \cite{bachmann2019multi}.

\section{Evaluation} \label{evaluation}






We conducted three experiments in simulation, each with a different pair of physical and virtual environments (see \autoref{subsec:env_layout}).
For each experiment, we compared our controller with two reactive controllers: an artificial potential function-based algorithm (APF) and steer-to-center (S2C).
The APF controller is implemented as described by Thomas et al. \cite{thomas2019general}.
We compared our method with APF because it is currently the state-of-the-art reactive RDW controller, having been shown to perform well in empty environments, environments with obstacles, and environments with multiple users \cite{thomas2019general, messinger2019effects, bachmann2019multi, dong2020dynamic}.
We chose to also compare our method with S2C because it is a commonly-used benchmark in the RDW controller literature.
Our implementation of S2C is the same as the one developed by Hodgson et al. \cite{hodgson2008redirected} since it has many improvements over the original S2C algorithm proposed by Razzaque \cite{razzaque2005redirected}.
We note that S2C is expected to perform very poorly in some of our environments (\autoref{subsec:env_layout}) due to obstacles near the center of the PE, and that it is fairer to compare ARC against APF in these environments, but we still evaluated S2C in these conditions for the sake of completeness.
The reset policy used by APF and S2C was the modified reset-to-center policy described in \cite{thomas2019general}.
We also informally tested a proof of concept VR implementation to evaluate ARC in real PE/VE pairs, for which we implemented ARC in the Unity 2019.4.8f1 game engine, and ran tests using an Oculus Quest head-mounted device.
The participant in the proof of concept was one of the authors.

There are existing controllers that either have similar features to our alignment-based controller or were tested in similar environments.
Zmuda et al.'s FORCE controller \cite{zmuda2013optimizing} makes use of the core assumption of alignment, that users will not walk into obstacles in the VE, to get performance gains.
Nescher et al.'s MPCRed controller \cite{nescher2014planning} uses information about the VE to inform the decisions about gain selection.
Although those controllers are similar to our controller in some aspects, we did not compare our work against them in this paper because they are predictive controllers, while our algorithm is purely reactive.
Since predictive and reactive controllers have fundamental differences by definition \cite{nilsson201815}, it would not be a fair comparison.

It should be noted that while it is unfair to compare ARC to predictive controllers, it may not be completely fair to compare with reactive controllers, either.
ARC is not predictive in the sense that it does not explicitly predict the user's future trajectory.
However, by computing the user's proximity, ARC does implicitly ``predict'' where a user will travel, since the algorithm assumes that users will avoid virtual obstacles.
ARC is reactive in the sense that all information used to set redirection gains is computed on a frame-by-frame basis, and no future or past information is used in the steering process.
Thus, one can consider ARC to fall somewhere between predictive and reactive controllers, which suggests that the traditional taxonomy of redirection controllers \cite{nilsson201815} may need to be updated to include newer algorithms.

\subsection{Performance Metrics}
\label{subsec:evaluation}
For each experiment, we compared the performance of the controllers using three quantitative performance metrics.
The metrics we used are:
\begin{itemize}
    \item \textbf{Number of resets:} The number of times the user collided with a physical obstacle. This is a standard metric in RDW literature.
    \item \textbf{Average distance walked between resets:} The average of the physical distance walked on a path before incurring a collision. 
    \item \textbf{Average alignment:} The average alignment $A(Q_t)$ for a path.
\end{itemize}
We also include qualitative evaluations showing the amount of space in the physical environment that was used, showing the effects of CR on controller performance, and showing the distribution of curvature gains applied by each controller.

The number of resets and the average distance walked between resets both provide a measurement of how many collisions the user incurs during locomotion.
The more collisions a user experiences, the shorter the distance between resets will be.
We also use heat maps 
of the user's location in the physical environment as a qualitative metric for the number of collisions.
When a user collides with an obstacle, they start a reset maneuver (\autoref{subsubsec:reset}) which involves turning in place where they stand.
This is manifested as a large amount of time spent in one spot, which will be highlighted on the heat map.
To measure and compare the intensity of gains applied by each controller, we computed the average curvature gain for each path walked, and present histograms of the frequency of each average curvature gain across all paths.
This metric is the same as the average steering rate metric that commonly appears in RDW literature \cite{strauss2020steering,bachmann2019multi,lee2019real}.
The average alignment metric is used to show that our algorithm does indeed optimize for a low alignment score, and that other algorithms do not.


\subsection{Simulated Framework}
Properly evaluating an RDW controller requires testing the controller on a large number of paths, ideally in varied environments.
It is common to use simulations to evaluate RDW controllers to avoid the high cost of running user studies \cite{thomas2019general, thomas2020reactive, strauss2020steering, bachmann2019multi, messinger2019effects, dong2020dynamic, chang2019redirection, lee2019real, thomas2020towards}.
Our simulated experiments were conducted on a computer with an AMD Ryzen 7 3700X 8-Core processor (3.60 GHz), 16 GB of RAM, a GeForce RTX 2080 SUPER GPU, and 64-bit Windows 10 OS. 

In an effort to make it easier to compare out work to prior research, our simulated user representation is similar to the one used by Thomas et al. \cite{thomas2019general}.
The user was represented as a circle with radius $0.5m$.
If the boundary of this circle came within $0.2m$ of an obstacle, this was counted as a collision and a reset was initiated.
The user's walking velocity was $1m/s$, and their angular velocity was $90^\circ/s$.
The path model used to generate user trajectories is the same as the one developed by Azmandian et al. \cite{azmandian2015physical}.
In this model, a waypoint is generated at a random distance ranging from $2m$ to $6m$ away from the previous waypoint.
The waypoint was placed at a random angle between $\pi$ and $-\pi$ relative to the previous waypoint.
To follow a series of waypoints, the user turned to face the next waypoint, and then walked in a straight line towards it.
Our simulation ran with a timestep of $0.05$.
To compute distances to obstacles, we represent the PE, VE, and obstacles as polygons (sets of vertices).

\subsection{Environment Layouts}
\label{subsec:env_layout}
Each of our three simulation experiments had a unique pair of physical and virtual environment configurations.
Diagrams for each environment are shown in \autoref{fig:environments}.
Exact coordinate data of the environment layouts is included in the supplementary materials.

\textbf{Environment A} includes an empty $10m \times 10m$ physical environment and an empty $10m \times 10m$ virtual environment.
This simple environment is used as a sanity check and to show that our algorithm can guide users on collision-free paths if perfect alignment is achievable.
The CR of Environment A is $1$.

\textbf{Environment B} is a moderately complex environment.
The physical environment is a $12m \times 12m$ physical room with $2m$-wide corridors.
These corridors are created by four $3m \times 3m$ square obstacles placed in the four quadrants of the room.
The virtual space for Environment B is a $17m \times 12m$ room with $2m$-wide corridors, created by six $3m \times 3m$ square obstacles.
Environment B was used to show that ARC can handle environment pairs that have locally similar features (regular corridors of the same width) but globally different dimensions.
The CR of Environment B is $1.170$.

\textbf{Environment C} is a highly complex environment.
The physical environment is a $10m \times 10m$ physical room with three rectangular obstacles.
In the center of the space is a $2m \times 4m$ obstacle.
The bottom-left corner of the room features a $2m \times 2m$ square obstacle.
Along the top boundary of the room is a $1m \times 4m$ obstacle.
This PE was designed to represent a plausible layout for a room in a house (such as a living room).
The virtual space used in Environment C is a $20m \times 20m$ room with regular and irregular polygonal obstacles scattered throughout the room.
Environment C was used to show that our algorithm is able to steer users through environments that are different in both local and global features.
The CR of Environment C is $1.625$.

We also used two different PE/VE pairs in our proof of concept implementation.
The first environment pair included a roughly $3.8m \times 6.9m$ PE and a roughly $5.65m \times 8.7m$ VE.
The PE was empty, and the VE had a roughly $1.7m \times 2.3m$ obstacle in the northeast corner of the room.
The second PE/VE pair featured a roughly $4.87m \times 7.62m$ PE and the same virtual room from the first environment pair.
The PE had a $1.1m \times 1.5m$ obstacle along the west wall, and the VE had a $1.7m \times 2.3m$ obstacle on the east wall, and a $1.7m \times 1.8m$ obstacle along the west wall.

\begin{figure}[t]
    \centering
    \includegraphics[width=.45\textwidth]{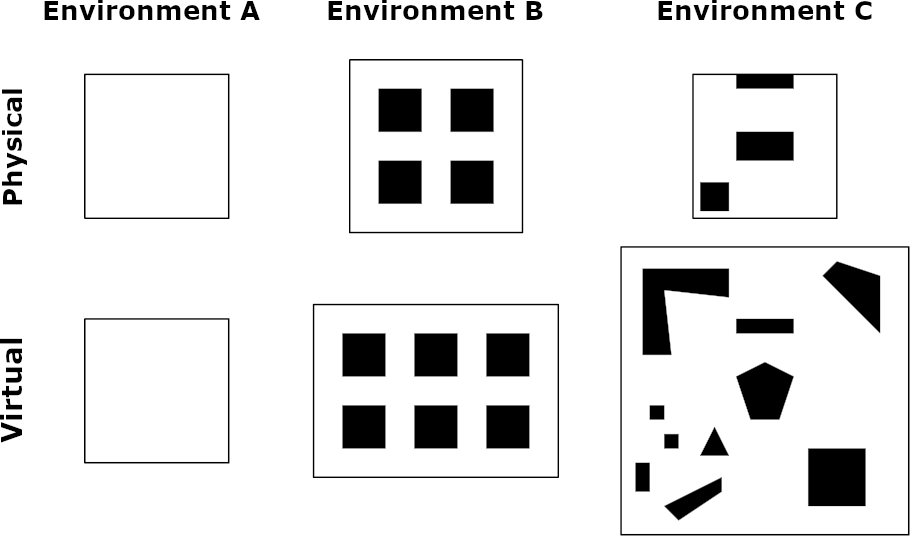}
    \caption{Diagrams of the physical and virtual environment pairs tested in our experiments. 
    Full descriptions of the environment layouts can be found in the supplementary material.}
    \label{fig:environments}
\end{figure}

\subsection{Experiment Design}
For each experiment, we generated 100 random, collision-free virtual paths with 100 waypoints.
The user travelled along each path three times, using either S2C, APF, or ARC for redirection.
Note that the \textit{same} 100 paths were used for each redirection controller within a particular environment.
Although the virtual paths were random, they all had the same starting location and direction within an environment.
In Experiment 1, the virtual user started in the center of Environment A, facing north.
In Experiment 2, the virtual user started in the center of Environment B, facing north.
In Experiment 3, the virtual user started $3.5m$ below the center of Environment C (south of the pentagon), facing north.
For every path, the physical user had a random starting location, but their heading direction matched that of the virtual user's at the start of the path.
The physical starting location for a given path was the same regardless of the controller being evaluated.
Having the user start in a random physical location increases the dissimilarity between the user's physical and virtual states, which makes it harder for a redirection controller to avoid collisions.
We used these random starting positions to show that ARC is still able to achieve a low number of collisions, even in this difficult setting.



\section{Results} \label{results}
We performed evaluations of the performance of S2C, APF, and ARC using three quantitative metrics and three qualitative metrics (\autoref{subsec:evaluation}).
Each metric was computed for all 100 paths in each condition.
RDW controller algorithms can sometimes encounter ``unlucky'' virtual paths that make it particularly difficult to avoid collisions, and the user will end up incurring many collisions in a short time frame.
To make our comparisons robust to these unlucky paths, outliers in the data that were $1.5$ times larger than the interquartile range of the data were replaced with the median of the data.
We evaluated the normality of the data using visual inspection of Q-Q plots and histograms as well as measures of the distributions' skew and kurtosis.
Homoscedasticity was evaluated using Levene's test.
For each metric measured, the assumption of normality or homoscedasticity was violated, so we conducted all of our tests using a robust one-way repeated measures $20\%$ trimmed means ANOVA (using the WRS2 package for R).
Pairwise post-hoc comparisons were computed using linear contrasts.

Due to the large sample size in our experiments, we report $95\%$ confidence interval instead of \textit{p}-values.
As the sample size grows, statistical tests become sensitive to small differences between samples, and the \textit{p}-value becomes unreliable as it approaches $0$.
Confidence intervals, however, become narrower as the sample size grows, which means they are able to scale with the sample size and are still reliable for experiments with large samples.
Furthermore, reporting confidence intervals allows for easier comparisons with future work if they also report confidence intervals, since the interval provides numerical bounds on the differences between conditions \cite{lin2013research}.

The results are shown in \autoref{tab:results}, with discussions in subsequent sections.
Since the distance walked between resets depends on the number of resets, we do not present the full analyses of the distance walked.
Instead, we only include the average differences in distances walked (see \autoref{tab:results}).
Full analyses on the distance walked between resets can be found in the supplementary materials.
Most of the discussion is limited to comparing ARC to APF, since it is already well-established that APF outperforms S2C \cite{thomas2019general, bachmann2019multi}.
A confidence interval that does not contain 0 in its range [CI lower, CI upper] is considered to be significant.
Thus, we found significant differences between all groups in our post-hoc tests.
The common pattern seen in the results is that ARC outperforms APF and S2C, while APF outperforms S2C.
One condition for which this was not the case is the alignment metric in Environment C.
Boxplots, heat maps, and frequency plots that visualize the results and redirection data are also included in the supplementary materials.


\linespread{1} 
\begin{table*}[ht]
    \centering
    \begin{tabular}{l|ccc|ccc|ccc}
    \multicolumn{10}{c}{\textbf{Environment A}}    \\ 
     & \multicolumn{3}{c}{Number of resets} & \multicolumn{3}{c}{Distance walked between resets} & \multicolumn{3}{c}{Average alignment score}   \\ \hline
        Redirection Controller & $\hat{\psi}$ & CI lower & CI upper & $\hat{\psi}$ & CI lower & CI upper & $\hat{\psi}$ & CI lower & CI upper  \\ \hline \hline
        S2C \cite{hodgson2013comparing} vs. ARC & 16.983 & 14.066 & 19.901 & -14.163 & -17.857 & -10.470 & 1.290 & 1.266 & 1.312  \\ \hline
        APF \cite{thomas2019general} vs ARC & 5.750 & 2.686 & 8.814 & -8.809 & -12.967 & -4.651 & 0.492 & 0.465 & 0.519  \\ \hline
        S2C \cite{hodgson2013comparing} vs 
        APF \cite{thomas2019general} & 11.617 & 9.148 & 14.086 & -5.822 & -6.808 & -4.836 & 0.801 & 0.789 & 0.813  \\
    \multicolumn{10}{c}{}    \\ 
    \multicolumn{10}{c}{\textbf{Environment B}}    \\ 
     & \multicolumn{3}{c}{Number of resets} & \multicolumn{3}{c}{Distance walked between resets} & \multicolumn{3}{c}{Average alignment score}   \\ \hline
        Redirection Controller & $\hat{\psi}$ & CI lower & CI upper & $\hat{\psi}$ & CI lower & CI upper & $\hat{\psi}$ & CI lower & CI upper  \\ \hline \hline
        S2C \cite{hodgson2013comparing} vs. ARC & 46.867 & 32.784 & 60.950 & -0.507 & -0.630 & -0.385 & 0.808 & 0.778 & 0.838  \\ \hline
        APF \cite{thomas2019general} vs ARC & 11.317 & 3.449 & 19.184 & -0.130 & -0.237 & -0.023 & 0.761 & 0.747 & 0.774  \\ \hline
        S2C \cite{hodgson2013comparing} vs 
        APF \cite{thomas2019general} & 36.408 & 23.560 & 49.256 & -0.382 & -0.502 & -0.263 & 0.033 & 0.002 & 0.065  \\
    \multicolumn{10}{c}{}    \\ 
    \multicolumn{10}{c}{\textbf{Environment C}}    \\ 
     & \multicolumn{3}{c}{Number of resets} & \multicolumn{3}{c}{Distance walked between resets} & \multicolumn{3}{c}{Average alignment score}   \\ \hline
        Redirection Controller & $\hat{\psi}$ & CI lower & CI upper & $\hat{\psi}$ & CI lower & CI upper & $\hat{\psi}$ & CI lower & CI upper  \\ \hline \hline
        S2C \cite{hodgson2013comparing} vs. ARC & 2139.983 & 2062.215 & 2217.752 & -3.719 & -3.792 & -3.645 & -1.036 & -1.060 & -1.012  \\ \hline
        APF \cite{thomas2019general} vs ARC & 137.108 & 124.065 & 150.152 & -2.296 & -2.420 & -2.171 & -0.109 & -0.131 & -0.087  \\ \hline
        S2C \cite{hodgson2013comparing} vs 
        APF \cite{thomas2019general} & 2005.500 & 1928.841 & 2082.159 & -1.414 & -1.488 & -1.339 & -0.928 & -0.961 & -0.894  \\
    \end{tabular}
    \caption{The results of pairwise post-hoc comparisons between controllers, computed using linear contrasts and reported using confidence intervals due to the large sample size \cite{lin2013research}. For each metric, $\hat{\psi}$ is the difference in estimated means between the two groups (estimate of the true mean). CI lower is the lower bound of the confidence interval on this difference, and CI upper is the upper bound. Narrower intervals indicate a more precise estimate of the true mean. We can interpret a cell as the estimated difference between the group means ($\hat{\psi}$), and CI lower and CI upper to represent that on $95\%$ of samples, the true difference in means between the groups will fall in the range $[\hat{\psi} - \text{CI lower}, \hat{\psi} + \text{CI upper}]$. For a given row that compares Algorithm X vs. Algorithm Y, a positive $\hat{\psi}$ value indicates that Algorithm X scored more than Algorithm Y by that $\hat{\psi}$, while negative a value indicates that Algorithm X scored lower than Algorithm Y by that $\hat{\psi}$, bounded by CI lower and CI upper.}
    \label{tab:results}
\end{table*}
\linespread{0.915} 


\subsection{Experiment 1 (Environment A)} \label{sec:exp1}
\label{results:exp1}
\subsubsection{Number of resets}
The robust repeated-measures ANOVA revealed a significant effect of redirection controller on the number of resets in Environment A $F(1.68,98.95)=126.1711, p<.0001$.
ARC outperformed both S2C and APF because it was usually able to steer the user on the same virtual paths as S2C and APF, but with fewer collisions.
We noticed that ARC sometimes performed worse than APF.
However, ARC was also sometimes able to achieve perfect alignment and steer users along paths with no collisions, which APF and S2C were not able to do.

In this experiment, APF outperformed S2C.
However, in the implementation of APF by Thomas et al. \cite{thomas2019general}, they did not find significant differences in the number of resets between APF and S2C in Environment A.
Since APF steers the user towards the center of the room, it is expected that APF and S2C will have similar results, as they did in \cite{thomas2019general}.
The difference in performance between APF and S2C is possibly explained by the stronger curvature gains applied by APF, since our implementation of S2C \cite{hodgson2013comparing} includes gain smoothing, whereby curvature gains transition gradually between values instead of instantly applying the strongest gain, as is done in APF \cite{thomas2019general}.



\subsubsection{Average alignment}
The robust trimmed means ANOVA revealed a significant effect of steering controller on the user's average alignment $F(1.39,82.19)=10870.26,p<.0001$.
The user's average alignment score was usually lower when they were redirected with ARC than when they were redirected with either APF or S2C.
Neither APF nor S2C is designed using concepts of alignment, so they should not be expected to achieve higher alignment scores than ARC does.
We also observed that user's alignment was consistently low regardless of the path, which signals the reliability of ARC in making the state at least close to aligned.


\subsubsection{Qualitative Evaluations}
For all conditions, the user spent most of their time near the center of the environment.
This is what S2C is designed to do, and APF reduces to S2C in empty environments, so this is no surprise.
One interesting thing to note, however, is that when the user is steered using ARC, they spend less time at the center of the room than with S2C and APF.


We observed differences in the average curvature gain applied by the controllers.
The implementation of APF we used \cite{thomas2019general} always applies maximum curvature gain.
However, S2C and ARC do not always apply the maximum curvature gain.
The average curvature gain for S2C is always above $6^\circ/s$, which is still fairly high considering the perceptual limit is $\approx7.6^\circ/s$.
On the other hand, ARC is able to apply curvature gains much lower than the perceptual limit and with high consistency, mostly ranging from $3^\circ/s$ to $6^\circ/s$.
All of the average gains applied by ARC are below $7^\circ/s$.

\subsection{Experiment 2 (Environment B)} \label{sec:exp2}
\subsubsection{Number of resets}
The robust ANOVA revealed a significant effect of controller on the number of resets $F(1.6,94.42)=56.0129,p<.0001$.
ARC and APF have somewhat similar performances, although ARC still resulted in significantly fewer resets than APF.
Furthermore, the interquartile range for the number of resets is lower for ARC than it is for APF, supporting the notion that ARC delivers a consistent locomotion experience that is robust to different virtual paths.

\subsubsection{Average alignment}
A significant effect of redirection controller on the user's average alignment was found $F(1.44,84.85)=3484.467,p<.0001$.
The same trend seen in Environment A for the average alignment score is also seen in Environment B.
ARC achieves a noticeably lower alignment score, which shows that ARC is able to successfully steer the user to a more aligned state.


\subsubsection{Qualitative Evaluations}
The physical position data showed differences between algorithms in where they steered the user.
S2C and APF have very similar heat maps since both algorithms steer the user towards the center of the space.
Interestingly, the user is able to visit most areas of the room using ARC, but there is a tendency to keep the user in the upper-left corner of the room.
It is possible that ARC is getting stuck between obstacles.
If that is the case, however, we would expect that the user gets stuck uniformly across the room due to their random starting locations, rather than getting stuck in one corner.


The average curvature gains showed a similar pattern as they did in Environment A.
S2C and ARC apply weaker curvature gains than APF.
One difference between Environment A and B is that the distribution of gains applied by ARC in Environment B is much smaller than it was in Environment A, which is likely because ARC was not able to achieve perfect alignment and thus was not able to apply small gains while also lowering the user's alignment score.

\subsection{Experiment 3 (Environment C)} \label{sec:exp3}
\subsubsection{Number of resets}
There was a significant effect of controller on the number of resets $F(1.04,61.34)=4186.948,p<.0001$.
ARC performs dramatically better than APF and S2C, with a much smaller spread in the number of collisions.
Paths steered by APF all have at least as many collisions as paths steered by ARC and are sometimes more than twice as bad as the worst path for ARC.

\subsubsection{Average alignment}
We found a significant effect of steering controller on the user's average alignment $F(1.04,61.34)=4186.948,p<.0001$.
An interesting pattern seen in the alignment scores for Environment C is that S2C scores the best (lowest) average alignment of all the controllers, and ARC has the highest alignment score.
This is surprising because neither S2C nor APF is designed to work based on alignment, but ARC is.
Even though ARC has the worst average alignment score of all the controllers in Environment C, it undoubtedly has the best performance in terms of number of collisions and physical distance travelled between resets.
This disagreement in the metrics suggests that the alignment metric we used in this study may not be a good representation of an alignment-based controller's ability to keep the system aligned and avoid collisions.
We stress that we \textit{do not} believe this means the alignment-based methods ARC uses to steer the user are flawed, since all other results in this section indicate that ARC \textit{does} work well compared to other controllers.
It may simply be the case that reporting the averaged sum of the user's forward and lateral alignment is not a good way to measure a controller's alignment capabilities since the results from Environment C show that it is possible for a controller that does not use alignment to have a better alignment score.
Another possible explanation for the differences in average alignment seen for Environment C is that the amount of distances we sample to compute distance to obstacles ($k=3$, see \autoref{alignment-definition}) may not be enough to accurately capture proximity in this environment.
Since this paper is only the second to formally study concepts of alignment, our alignment metric can likely be improved.


\subsubsection{Qualitative Evaluations}
Upon observing the physical position heat maps, we noticed that ARC utilizes more of the PE for navigation than do APF or S2C, but there is still a bias towards the leftmost region of the room.
Visual inspection of the random starting positions in the PE confirmed that the bias was not due to the starting position, so more work should be done to get a better understanding of the biases of ARC.

We inspected the frequency plot of gains applied by each controller across all paths in Environment C.
All gains applied by both ARC and S2C are lower than those of APF, with ARC applying the lowest gains of all, while applying stronger gains less frequently than S2C.

\subsection{Proof of Concept Implementation}
We implemented ARC in a VR system using an Oculus Quest and the Unity 2019.4.8f1 game engine.
Our proof of concept implementation shows that ARC works as intended in real VR systems, but we note that a full user study should be conducted before drawing more conclusions about ARC in VR systems.
In both environments that we tested, the user's virtual position started centered along the south wall of the VE, and the user was instructed to walk in a straight line forward in the VE.

The first PE/VE pair was intentionally designed to be fairly simple in order to verify that ARC steers the user as it should.
The user's physical position started in the southeast corner of the PE.
Once the user started walking forward, they were steered to the left, away from the east wall of the PE, using curvature gains.
The user was steered away from the east wall in order to improve their alignment with the virtual state, since the virtual user was not beside any walls.

In the second PE/VE pair, the user's physical location started in the southwest corner of the physical room.
Similarly to the first environment, the user was steered away from the nearby physical wall.
Once their virtual position was between the two obstacles in the VE, ARC continued to steer the user to the right with curvature gains, in an effort to minimize the misalignment between the user's physical and virtual states.
When the user walked past the virtual obstacle on the left, ARC steered the user slightly towards the left to improve their alignment.

The proof of concept implementation shows that our algorithm is able to run in real time without interfering with the user's ability to travel on an intended virtual path, which is a crucial requirement for redirection controllers.
Additionally, ARC is simple enough such that it can be used on each frame without negatively impacting the frame rate of the system.
Recordings of the user walking in each PE/VE pair can be found in the supplementary materials video.


\section{Discussion} \label{analysis}

We found that a redirection controller based on alignment can be a very effective alternative to traditional controllers that always try to steer users away from physical obstacles.
Our novel alignment-based redirection controller, ARC, outperformed current state-of-the-art methods in all environments that we tested for all metrics except for average alignment in Environment C.
In addition to being able to deliver a locomotion experience with fewer collisions and further distances walked between collisions, ARC steers the user with curvature gains that are less intense than those applied by other controllers.
ARC achieves this high performance using just three distance calculations from the VE and three from the PE, which allows it to easily run in real time.
Using information from the VE has usually only been done by predictive controllers, but we were able to develop a reactive controller that leverages instantaneous information from the VE for large performance benefits, and does not require complex predictions about the user's behavior.

We also presented Complexity Ratio (CR), a new metric to measure the relative complexity of a pair of physical and virtual environments by describing the density of obstacles in the environments.
The relative complexity of environments is an important factor in a controller's ability to steer the user, but we are unaware of any RDW studies that have explicitly defined and discussed any notions of relative complexity and how it affects controllers' performance.
Our work presents the first step in this direction.
We showed that traditional controllers tend to perform worse as the difference in complexity between the PE and VE grows.
This agrees with prior observations that the shape of the environment affects a controller's performance \cite{azmandian2015physical, hodgson2013comparing}.

ARC comes with many advantages over traditional steering policies.
First, ARC decreases the likelihood that a user experiences simulator sickness due to strong redirection.
While the exact cause of simulator sickness is not known, one of the main theories is that simulator sickness arises when there is a conflict between visual, vestibular, and proprioceptive stimuli \cite{kolasinski1995simulator}.
RDW creates this exact perceptual conflict, so it is not uncommon for users to feel simulator sickness when being redirected.
Although we know it is safe to apply redirection within the perceptual thresholds, these thresholds will vary from user to user.
Thus, the we cannot assume that commonly purported threshold values will be suitable for all users.
By only applying redirection when the user is misaligned, and only applying gains at the intensity necessary to achieve alignment, ARC redirects the user less than a traditional controller does, which decreases the likelihood of simulator sickness and creates a more comfortable experience.

Steering by alignment provides passive haptics by enabling the user to interact with the physical environment \cite{thomas2020towards, thomas2020reactive}.
Passive haptics are physical objects that provide feedback to the user through their shape and texture \cite{lindeman1999hand}.
Passive haptics can significantly increase a user's feelings of presence and spatial knowledge transfer \cite{insko2001passive}.
Passive haptics and RDW have typically been considered mutually exclusive due to their conflicting requirements.
However, Kohli et al. \cite{kohli2005combining} demonstrated that it is possible to combine the two if we have the appropriate environment configurations.
Their demonstration was in a carefully crafted environment designed specifically to enable passive haptics.
Alignment can enable passive haptics in arbitrary environments, which may allow for more immersive experiences that combine comfortable locomotion through redirection with realistic sensations through passive haptics.
The efficacy of using alignment to combine passive haptics and RDW should be studied through formal user studies, since alignment currently does not consider the shape and orientation of obstacles, which are important factors for effective passive haptics \cite{kohli2005combining}.

\section{Conclusions and Future Work} \label{conclusion}

In this paper we presented ARC, a novel controller based on alignment.
Through extensive simulation-based experiments, we showed that our controller was able to outperform state-of-the-art algorithms in both simple and complex environments.
Furthermore, our algorithm applied redirection gains at a lower intensity than other controllers, which reduces the chances of inducing simulator sickness and improves the usability of RDW systems for people with low RDW perceptual thresholds.
We also formalized the notion of relative environment complexity between the physical and virtual environments, which to the best of our knowledge had not yet been done.
To this end, we introduced Complexity Ratio (CR), a novel metric to measure the difference in navigation complexity between the physical and virtual environments, and showed how CR influences controller efficacy.

There are many avenues for future work.
The heuristics that ARC uses are fairly simple, so it is likely that a more complex algorithm will yield a better performance.
For example, a finer approximation of the user state using more distance samples may yield better results.
Additional work should also be done to get a better understanding of the biases that ARC exhibited, so we can better predict how a controller will perform in an environment.
Extending ARC to dynamic scenes with moving obstacles or multiple users is also an interesting area for future work.
Furthermore, ARC should also be evaluated with full user studies now that we know that alignment can be an effective method for redirection.
Future work should also investigate ways to use concepts of alignment to combine passive haptics with redirected walking.


It is currently quite difficult to compare controllers from different researchers without implementing them oneself, since experiments are often conducted under very different conditions.
The ability to compare controllers may help the community to develop new controllers more effectively, since direct comparisons will highlight the strengths and weaknesses of controllers.
To enable comparisons between RDW controllers, work should be done to develop accurate performance metrics and standard benchmarks.
It is likely that development of good metrics and benchmarks will require a deep understanding of the complicated interactions between the PE, the VE, the virtual path, and the controller, since any good metrics and benchmarks will need to encapsulate these interactions.


\acknowledgments{
The authors wish to thank Tabitha Peck for her helpful comments on statistical analysis, and the reviewers for their insightful suggestions.
This work was supported in part by ARO under Grants W911NF1910069 and W911NF1910315, and in part by Intel.
}

\bibliographystyle{abbrv-doi}

\bibliography{references}

\newpage 

\renewcommand{\thesection}{\Alph{section}}
\setcounter{section}{0}
\section{Supplementary Materials} \label{sec:supplementary}

\subsection{Environment A: Analysis of Average Distance Walked Between Resets}
There was a significant effect of controller on the distance walked $F(1.09,64.09)=57.9766, p<.0001$.
A boxplot of the average distance walked between resets for each controller is shown in \autoref{fig:all_boxplots}, and the precise difference between controllers is shown in \autoref{tab:results}.
When navigating with ARC, the user was able to walk further without colliding with a physical obstacle when compared with APF and S2C.
The long upper whisker and the dots representing outlier paths indicate that for some paths, the simulated user was able to walk over $45m$ before colliding with an obstacle, which shows that ARC is able to deliver VR experiences with very few resets.

\subsection{Environment B: Analysis of Average Distance Walked Between Resets}
There was a significant effect of steering algorithm on the average distance walked between resets $F(1.9,112.05)=58.9188 ,p<.0001$.
A plot of the average distances walked between resets for all paths with all controllers is shown in \autoref{fig:all_boxplots}.
The results of post-hoc tests to determine the differences between controllers is shown in \autoref{tab:results}.
The boxplot for the physical distances walked in Environment B shows that ARC achieves a higher median distance than APF and S2C, but the largest average distances afforded by ARC are not as big as the longest distances walked with APF.
This suggests that APF may be more suited than ARC for navigation in environments with corridors, like Environment B, but additional studies should be conducted to confirm this.

\subsection{Environment C: Analysis of Average Distance Walked Between Resets}
\begin{figure*}[t]
    \centering
    \includegraphics[width=\linewidth]{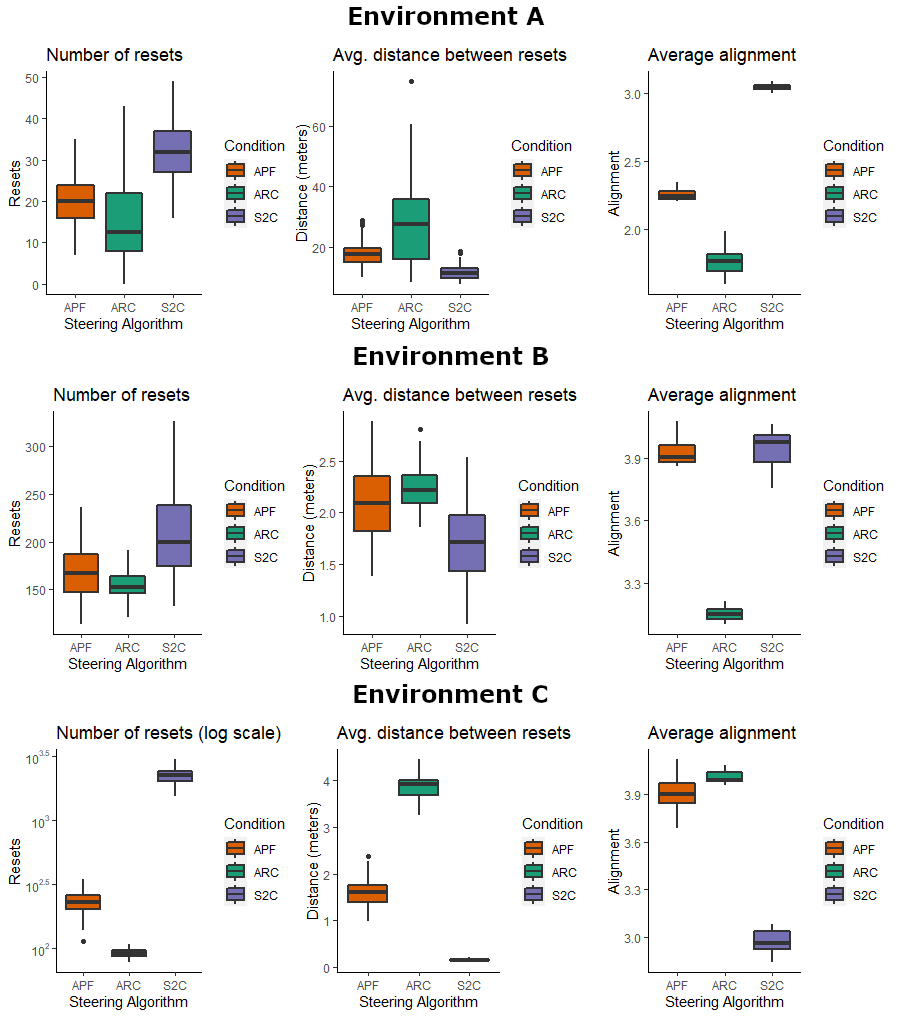}
    \caption{Boxplots of performance metrics for each controller in each environment. The boxplots show the median and IQR for the data. A significant difference was found between all algorithms in all environments. ARC outperformed APF and S2C for all metrics in all environments except for average alignment in Environment C.}
    \label{fig:all_boxplots}
\end{figure*}

A robust trimmed-means ANOVA revealed a significant effect of controller on the average physical distance walked by the user between resets $F(1.52,89.44)=5855.824,p<.0001$.
Boxplots showing the distributions of average distances walked between resets for all controllers are Environment C is shown in \autoref{fig:all_boxplots}, and the results from post-hoc significance tests are in \autoref{tab:results}.
ARC outperforms APF and S2C, and the results for ARC are more consistent than they are for APF, though there is not as dramatic a difference as there was for the number of resets.
The number of resets for Environment C shows that ARC performs much more consistently than APF, but the average distance between resets for Environment C, while it shows the same overall trend, suggests that the difference in consistency is not as large as it seemed from the number of resets, highlighting the importance of using multiple performance metrics.

\newpage
\subsection{Environment Layouts and Additional Figures}

\begin{figure}[t]
    \centering
    \includegraphics[width=.5\textwidth]{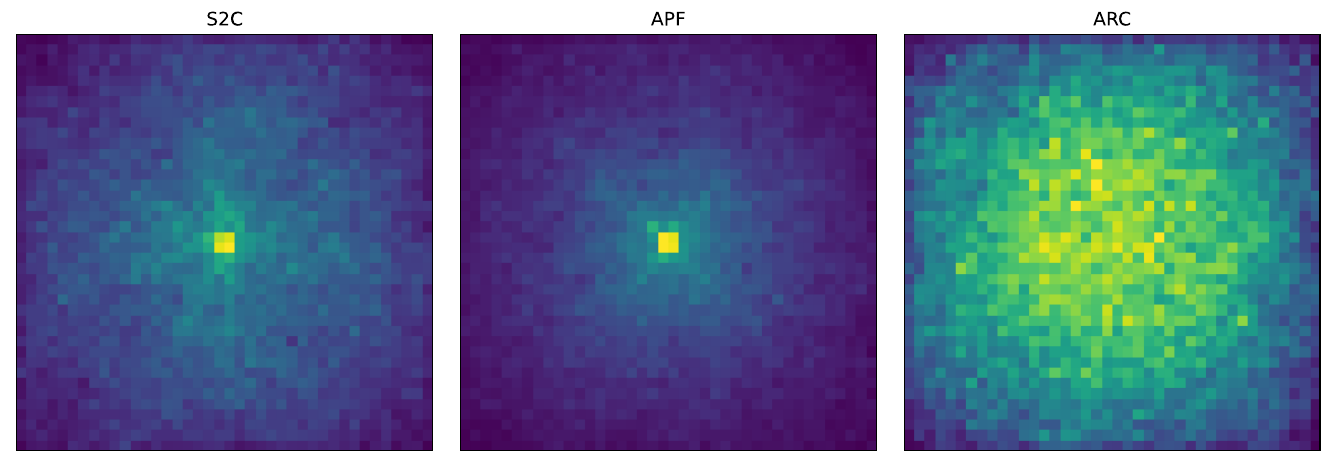}
    \caption{A heat map of the user's physical position across all paths for each controller in Environment A. Yellow tiles indicate the most time spent at that location, while purple tiles indicate the least amount of time. S2C and APF steer the user such that they spent the large majority of their time in the center of the room, while ARC allows the user to visit each region of the room more evenly.}
    \label{fig:envA_combined_heatmap}
\end{figure}

\begin{figure}[t]
    \centering
    \includegraphics[width=.5\textwidth]{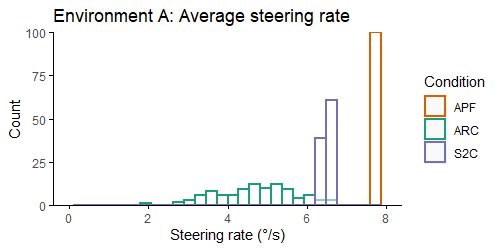}
    \caption{A histogram of the average curvature gain applied by each controller for each path in Environment A. The implementation of APF we used always applies the same gain, while S2C and ARC apply lower gains on average. S2C still applies gains fairly close to the perceptual threshold ($\approx 7.6^\circ /s$), but ARC is able to steer the user on paths with fewer collisions and significantly reduced curvature gains. Most of the gains applied by ARC fall in the $3^\circ/s - 5^\circ /s$ range, showing that ARC only applies the gains necessary to avoid collisions and maintain alignment.}
    \label{fig:envA_steering}
\end{figure}

\begin{figure}[t]
    \centering
    \includegraphics[width=.5\textwidth]{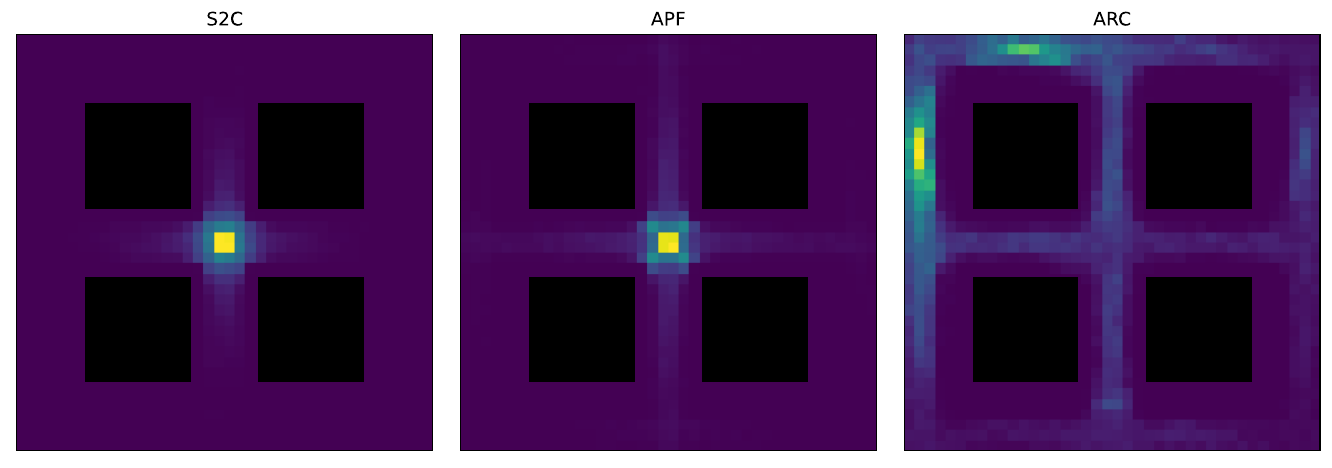}
    \caption{A heat map of the physical locations visited by the user in Environment B when steered with each controller. Yellow tiles indicate more visits to a region, while purple tiles indicate less time spent in a region. Obstacles are shown in black. S2C and APF keep the user concentrated near the center of the room since it is the most open space in all directions, while ARC is able to utilize more of the space and steer the user along all corridors in the room. ARC has some tendency to keep the user near the north wall of the room, which we suspect is due to the user getting stuck in between obstacles, but the exact cause is not clear.}
    \label{fig:envB_combined_heatmap}
\end{figure}

\begin{figure}[t]
    \centering
    \includegraphics[width=.5\textwidth]{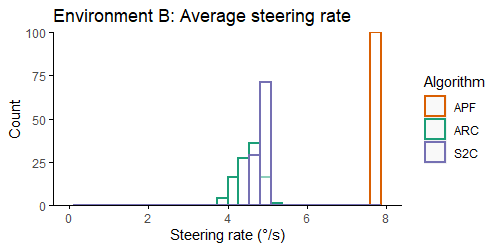}
    \caption{A histogram of the average curvature gain applied for each path with each controller in Environment B. As in Environment A, APF applies a constant curvature gain when the user is walking. S2C and ARC apply gains with an average in the range of $4^\circ/s - 6^\circ/s$, with ARC applying gains all gains at a lower intensity than about half of the gains applied by S2C. Note that the lowest gains applied by S2C are lower than those of ARC.}
    \label{fig:envB_steering}
\end{figure}

\begin{figure}[t]
    \centering
    \includegraphics[width=.5\textwidth]{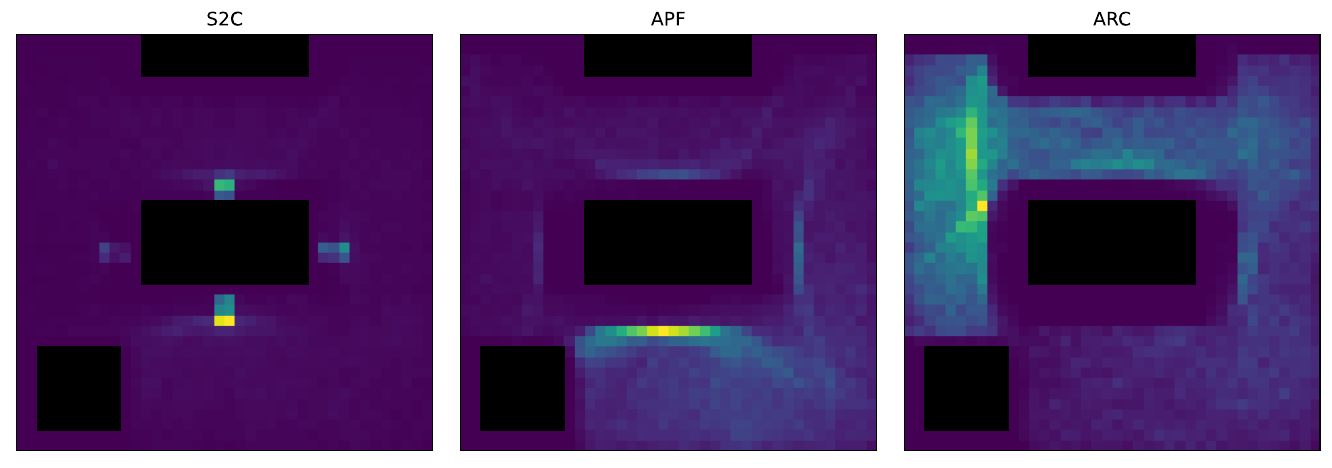}
    \caption{A heat map of the simulated user's location in the physical environment when exploring a virtual environment using three different redirection controllers. Yellow tiles represent a large amount of time spent in that region, and purple tiles represent a small amount of time spent in that region. The Alignment-based Redirection Controller (ARC) allows the user to utilize more of the physical space while exploring the virtual world compared to S2C and APF. This means that users spends less time being reset and more time walking through the physical environment, when steered with ARC than with S2C or APF. This is supported by the results for the number of collisions and distance walked.}
    \label{fig:envC_combined_heatmap}
\end{figure}

\begin{figure}[t]
    \centering
    \includegraphics[width=.5\textwidth]{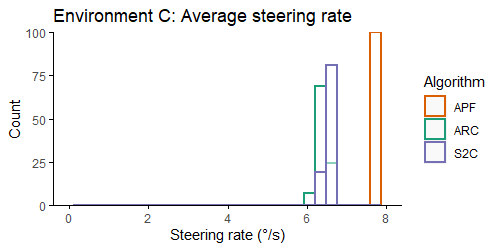}
    \caption{The average curvature gain applied by each controller for all paths in Environment C. The same trend as in Environment B is seen here, where APF has a higher steering rate than S2C and ARC. One difference between the steering rates in Environment B and C is that the gains applied by S2C and ARC are in a higher range ($6^\circ/s - 7^\circ/s$) in Environment C than they were in Environment B ($4^\circ/s - 6^\circ/s$).}
    \label{fig:envC_steering}
\end{figure}

\begin{figure}[t]
    \centering
    \includegraphics[width=.5\textwidth]{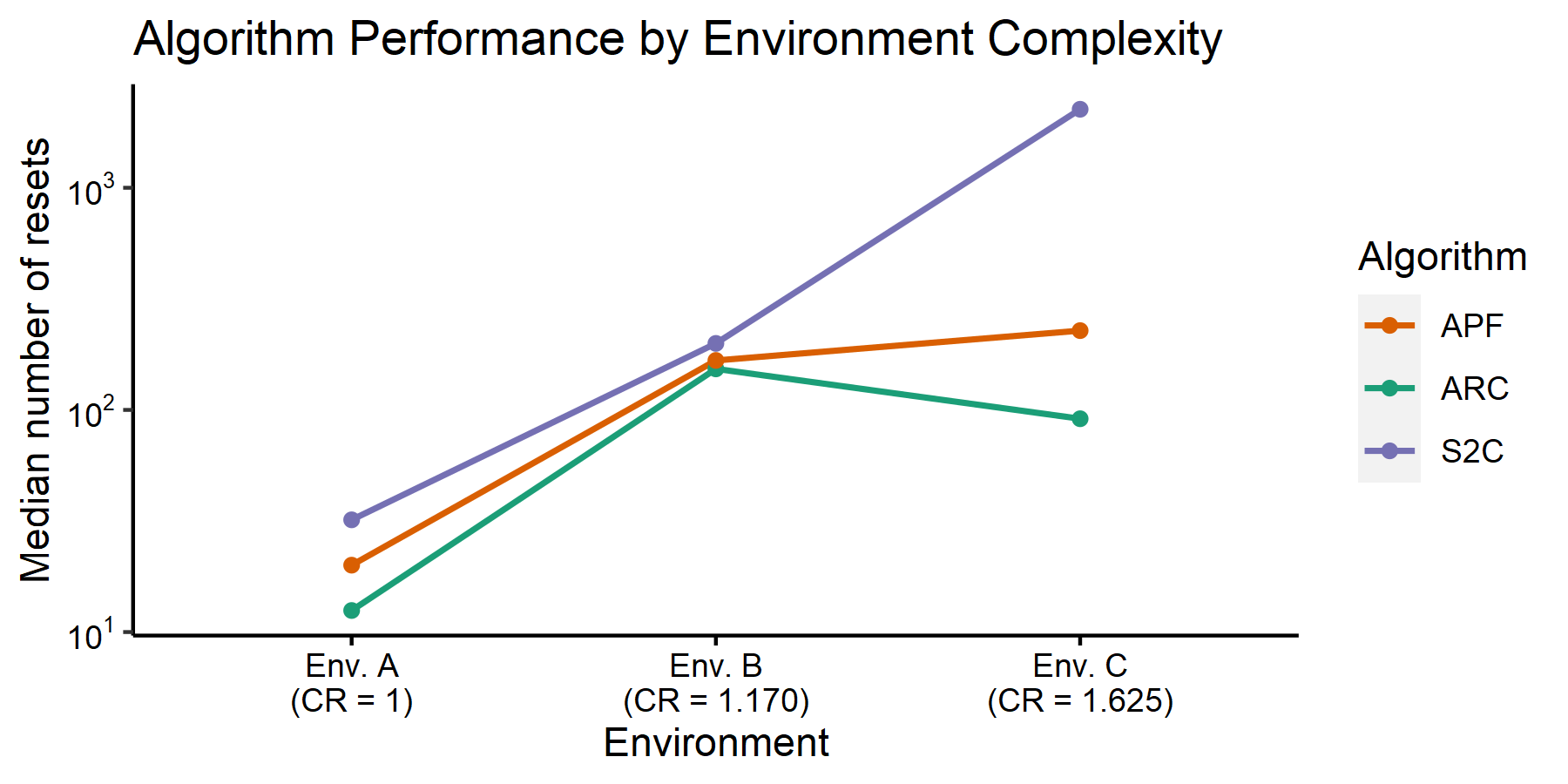}
    \caption{The relationship between the environment complexity and the number of resets incurred by a redirection controller. ARC consistently has a better performance than S2C and APF for all environment complexities. The performance difference between ARC and the other algorithms is quite large for environments A and C, but the difference decreases drastically for Environment B. It is not clear why Environment B causes the controllers to have a more similar performance, but it may be due to the relatively few pathing options afforded by the narrow hallways of Environment B. Environments A and C both include regions with a fairly large amount of open space, unlike Environment B (see \autoref{fig:environments}).}
    \label{fig:resets_by_CR}
\end{figure}

\begin{figure}[t]
    \centering
    \includegraphics[width=.5\textwidth]{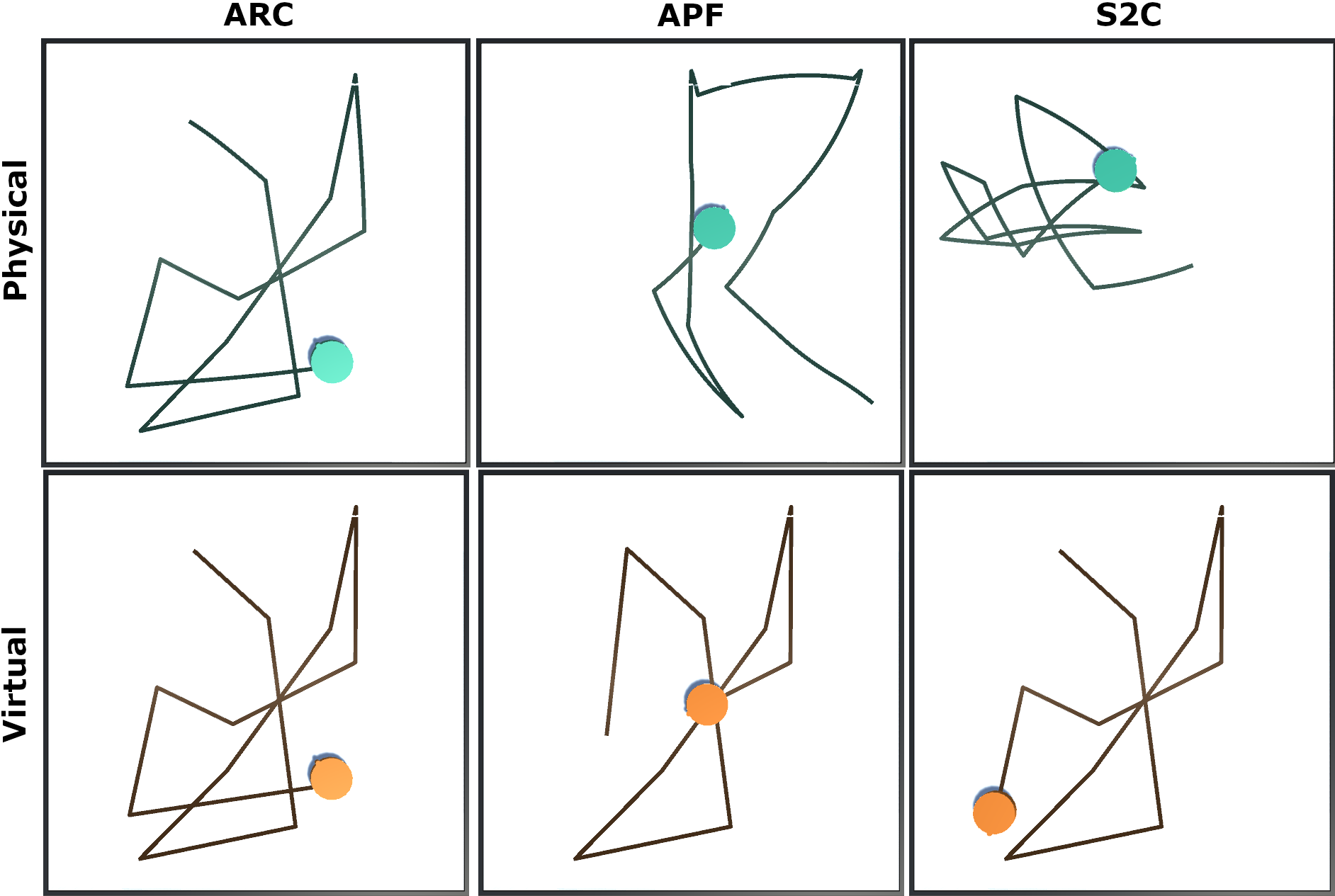}
    \caption{A screenshot of the user's state and recent path in Environment A for each controller. Each simulated user travelled on the same virtual path in this figure, and the screenshot was taken at the same time in the simulation. When steered with ARC, the system is able to achieve perfect alignment, and the user's physical state and recent path matches the virtual counterpart. APF and S2C are not able to achieve alignment, and their paths and states are very dissimilar to the virtual counterparts. The state of the virtual user is not the same across all conditions because the virtual user pauses while the physical user reorients after a collision, and each controller incurred a different number of collisions.}
    \label{fig:alignment_comparison}
\end{figure}

\begin{table}[hb]
    \centering
    \begin{tabular}{c|p{53mm}}
        \multicolumn{2}{c}{\textbf{Environment A (physical)}}    \\ 
        Boundary & $(-5, -5), (5, -5), (5, 5), (-5, 5)$ \\ 
        \multicolumn{2}{c}{}    \\ 
        \multicolumn{2}{c}{\textbf{Environment A (virtual)}}    \\ 
        Boundary & $(-5, -5), (5, -5), (5, 5), (-5, 5)$ \\ 
        \multicolumn{2}{c}{}    \\ 
        \multicolumn{2}{c}{\textbf{Environment B (physical)}}    \\ 
        Boundary & $(-6, -6), (6, -6), (6, 6), (-6, 6)$ \\ \hline
        Obstacle 1 & $(-4, -4), (-1, -4), (-1, -1), (-4, -1)$ \\ \hline
        Obstacle 2 & $(1, -4), (4, -4), (4, -1), (1, -1)$ \\ \hline
        Obstacle 3 & $(1, 1), (4, 1), (4, 4), (1, 4)$ \\ \hline
        Obstacle 4 & $(-4, 1), (-1, 1), (-1, 4), (-4, 4)$ \\ 
        \multicolumn{2}{c}{}    \\ 
        \multicolumn{2}{c}{\textbf{Environment B (virtual)}}    \\ 
        Boundary & $(-11, -6), (6, -6), (6, 6), (-11, 6)$ \\ \hline
        Obstacle 1 & $(-4, -4), (-1, -4), (-1, -1), (-4, -1)$ \\ \hline
        Obstacle 2 & $(1, -4), (4, -4), (4, -1), (1, -1)$ \\ \hline
        Obstacle 3 & $(1, 1), (4, 1), (4, 4), (1, 4)$ \\ \hline
        Obstacle 4 & $(-4, 1), (-1, 1), (-1, 4), (-4, 4)$ \\ \hline
        Obstacle 5 & $(-9, 1), (-6, 1), (-6, 4), (-9, 4)$ \\ \hline
        Obstacle 6 & $(-9, -4), (-6, -4), (-6, -1), (-9, -1)$ \\ 
        \multicolumn{2}{c}{}    \\ 
        \multicolumn{2}{c}{\textbf{Environment C (physical)}}    \\ 
        Boundary & $(-5, -5), (5, -5), (5, 5), (-5, 5)$ \\ \hline
        Obstacle 1 & $(-4.5, -4.5), (-2.5, -4.5), $ \newline $  (-2.5, -2.5), (-4.5, -2.5)$ \\ \hline
        Obstacle 2 & $(-2, -1), (2, -1), (2, 1), (-2, 1)$ \\ \hline
        Obstacle 3 & $(-2, 4), (2, 4), (2, 5), (-2, 5)$ \\ 
        \multicolumn{2}{c}{}    \\ 
        \multicolumn{2}{c}{\textbf{Environment C (virtual)}}    \\ 
        Boundary & $(10, -10), (10, 10), (-10, 10), (-10, -10)$ \\ \hline
        Obstacle 1 & $(-4.5, -4.5), (-2.5, -4.5), (-3.5, -2.5)$ \\ \hline
        Obstacle 2 & $(0, 2), (2, 1), (1, -2), (-1, -2), (-2, 1)$ \\ \hline
        Obstacle 3 & $ (-2,4), (2,4), (2,5), (-2,5) $ \\ \hline
        Obstacle 4 & $ (-8.5,8.5), (-8.5,2.5), (-6.5,2.5), $ \newline $ (-7,7), (-2.5,6.5), (-2.5,8.5) $ \\ \hline
        Obstacle 5 & $ (-8,-1), (-8,-2), (-7,-2), (-7,-1) $ \\ \hline
        Obstacle 6 & $ (-7,-3), (-7,-4), (-6,-4), (-6,-3) $ \\ \hline
        Obstacle 7 & $ (-9,-5), (-9,-7), (-8,-7), (-8,-5) $ \\ \hline
        Obstacle 8 & $ (-6,-9), (-3,-7), (-3,-6), (-7,-8) $ \\ \hline
        Obstacle 9 & $ (3,-4), (3,-8), (7,-8), (7,-4) $ \\ \hline
        Obstacle 10 & $ (5,9), (4,8), (8,4), (8,8) $ \\
    \end{tabular}
    \caption{Coordinates of vertices of boundaries and obstacles in each environment.}
    \label{tab:env_config}
\end{table}

\end{document}